\documentclass{article}  
\usepackage{jamaica04}
\usepackage{graphicx}
\usepackage{amssymb}
\usepackage{dcolumn}
\usepackage{amsmath}
\usepackage{supertabular}
\frompage{000} \topage{000}                                              
\def\rightmark{Identified particles at RHIC}
\def\leftmark{P. Sorensen}
 
\title{Identified particle measurements at RHIC: elucidating
  hadronization mechanisms for bulk partonic matter } \authors{
{Paul Sorensen %
}\\[2.812mm]
{\normalsize
Lawrence Berkeley National Laboratory, \\ 
94720 Berkeley, USA\\[0.2ex] 
}}
 
\abstract{ Measurements of identified particle momentum spectra at
  $\sqrt{s_{_{NN}}}=200$~GeV are reviewed. Emphasis is placed on the
  azimuthal dependence and the centrality dependence of hadron yields
  at intermediate transverse momentum ($1.5 < p_T < 5$ GeV/c). The
  first measurements of the fourth harmonic term ($v_4$) in the
  azimuthal variation of identified particle yields are shown. The
  recombination mechanism of hadron formation provides a consistent
  description of the dependence of these measurements on
  particle-type. }

 
\begin{document}
 
\maketitle
\setcounter{page}{1}

\section{Introduction}\label{intro}

Experimenters at the Relativistic Heavy Ion Collider (RHIC) have made
several unexpected observations. Perhaps most surprising are
measurements relating to baryon production in the intermediate
transverse momentum region ($1.5<p_T<5$
GeV/c)~\cite{Adler:2003kg,Adams:2003am}. At this $p_T$, while in
nucleon-nucleon collisions, one baryon is produced for every three
mesons (1:3), in Au+Au collisions baryons and mesons are created in
nearly equal proportion (1:1).  Although these initial observations
were considered a {\it puzzle}, they can easily be reconciled when
multi-parton mechanisms for hadron production such as recombination or
coalescence are considered~\cite{reco}. The azimuthal dependence of
production provides compelling support for this picture with baryon
and meson anisotropies scaling in a manner predicted by coalescence
models~\cite{Adams:2003am,sorensen}. Since these measurements were
made, no other picture has been presented that can account for this
apparent meson/baryon dependence. We note, however, that several of
the defining measurements --- particularly those for the more massive
mesons --- are either preliminary or statistically
limited~\cite{msb}. Data recently taken at RHIC will allow
experimentalists to measure identified particle distributions to
higher $p_T$ and test multi-parton hadronization predictions with
greater precision.

We use the ratio of yields from central and peripheral collisions, $R_{CP}$,
scaled by the number binary collisions $\mathrm{N_{bin}}$:
\begin{equation}
R_{CP}(p_T) = \frac{\left [dN/\left (\mathrm{N_{bin}}dp_T \right )
\right ]^{central}}{\left [dN/\left (\mathrm{N_{bin}}dp_T \right )
\right ]^{peripheral}},
\end{equation}
to study the evolution of a particle's $p_T$ spectra with collision
centrality. In Au+Au collisions at $\sqrt{s_{_{NN}}}=200$~GeV,
$R_{CP}(p_T>6~\mathrm{GeV/c}) \approx 0.2$~\cite{highpt}, indicating
that hadron yields in central collisions at this $p_T$ are a factor of
five smaller than expected from scaling of peripheral collisions. To
study production with respect to the collision orientation ({\it
  i.e. the reaction-plane direction}), the azimuthal component of a
particle's distribution is expanded as a Fourier series, $dN/d\phi
\propto 1 + \sideset{}{_n}\sum\nolimits 2v_n\cos n\phi$, where $\phi$
is the azimuth angle of the particle momentum with respect to the
reaction-plane direction~\cite{flow}. Owing to the elliptic shape of
the reaction volume in off-axis collisions, the second term $v_2$ is
the largest and most studied of the coefficients.

\section{Intermediate-$p_T$ defined}

Based on the yields and azimuthal anisotropies of identified particles
in relativistic nuclear collisions, three $p_T$ regions with distinct
characteristics have been delineated. The low-$p_T$ region is
compellingly characterized by a mass dependence of $v_2$ that
signifies the development of a collective
velocity~\cite{sorensen,pidv2}. At intermediate-$p_T$ a strong
number-of-constituent-quark (NCQ) dependence dominates both $v_2$ and
$R_{CP}$. At $p_T > 5$ GeV/c the lack of a species dependence in
$R_{CP}$ indicates that baryon and meson production increases with
centrality at similar rates~\cite{Adams:2003am}.

\begin{figure}[htpb]
\resizebox{.32\textwidth}{!}{\includegraphics{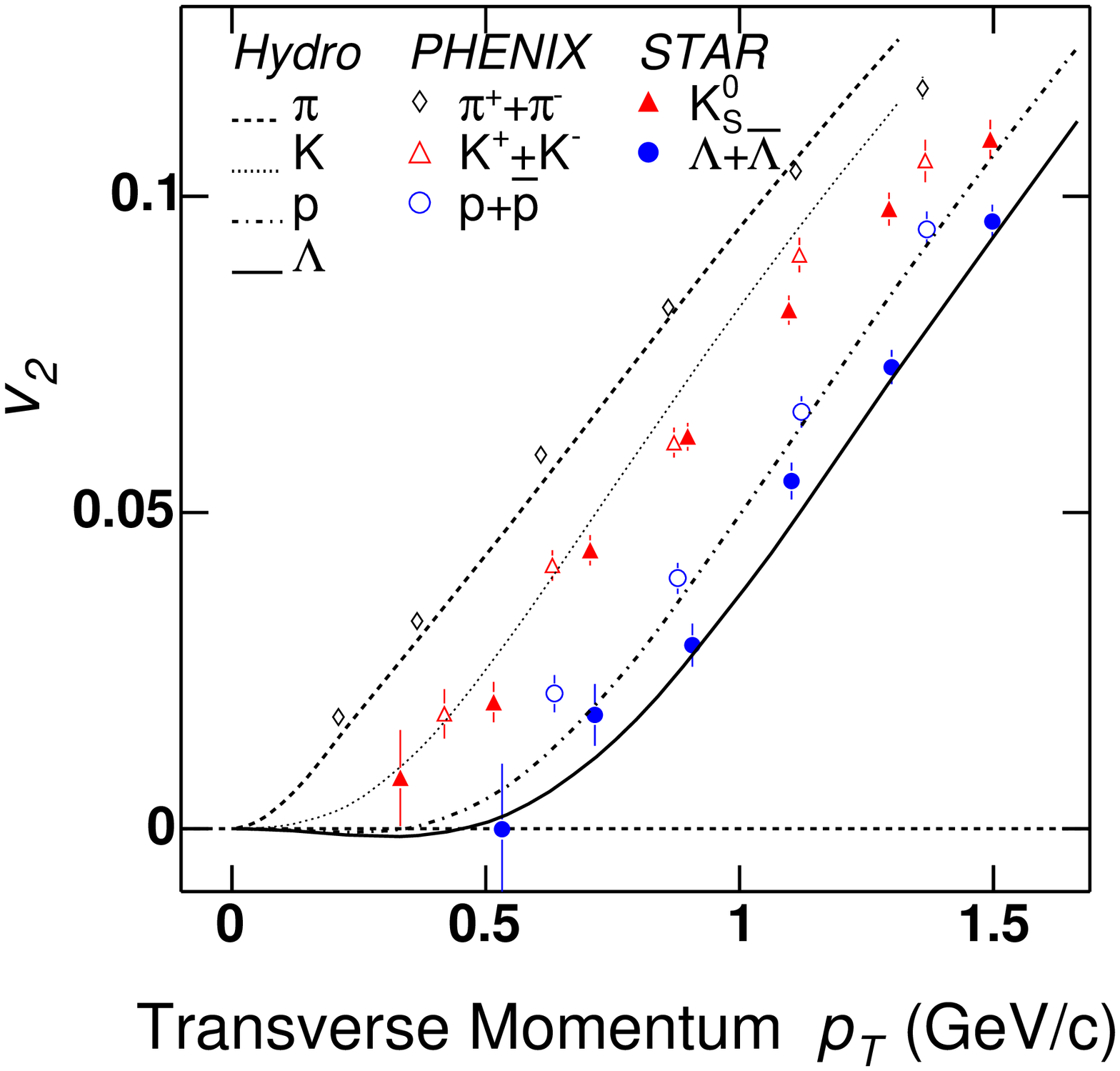}}
\resizebox{.32\textwidth}{!}{\includegraphics{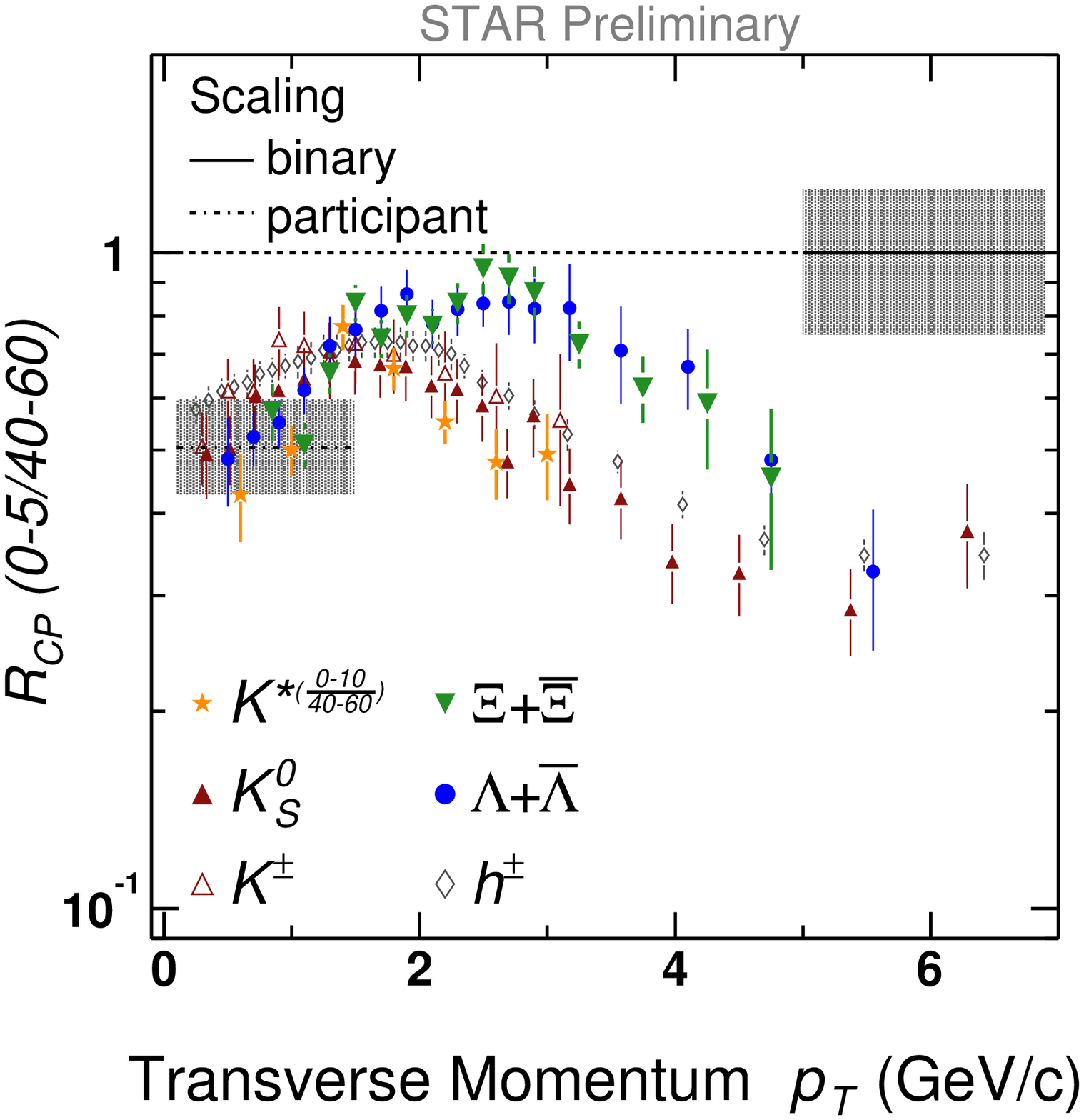}}
\resizebox{.32\textwidth}{!}{\includegraphics{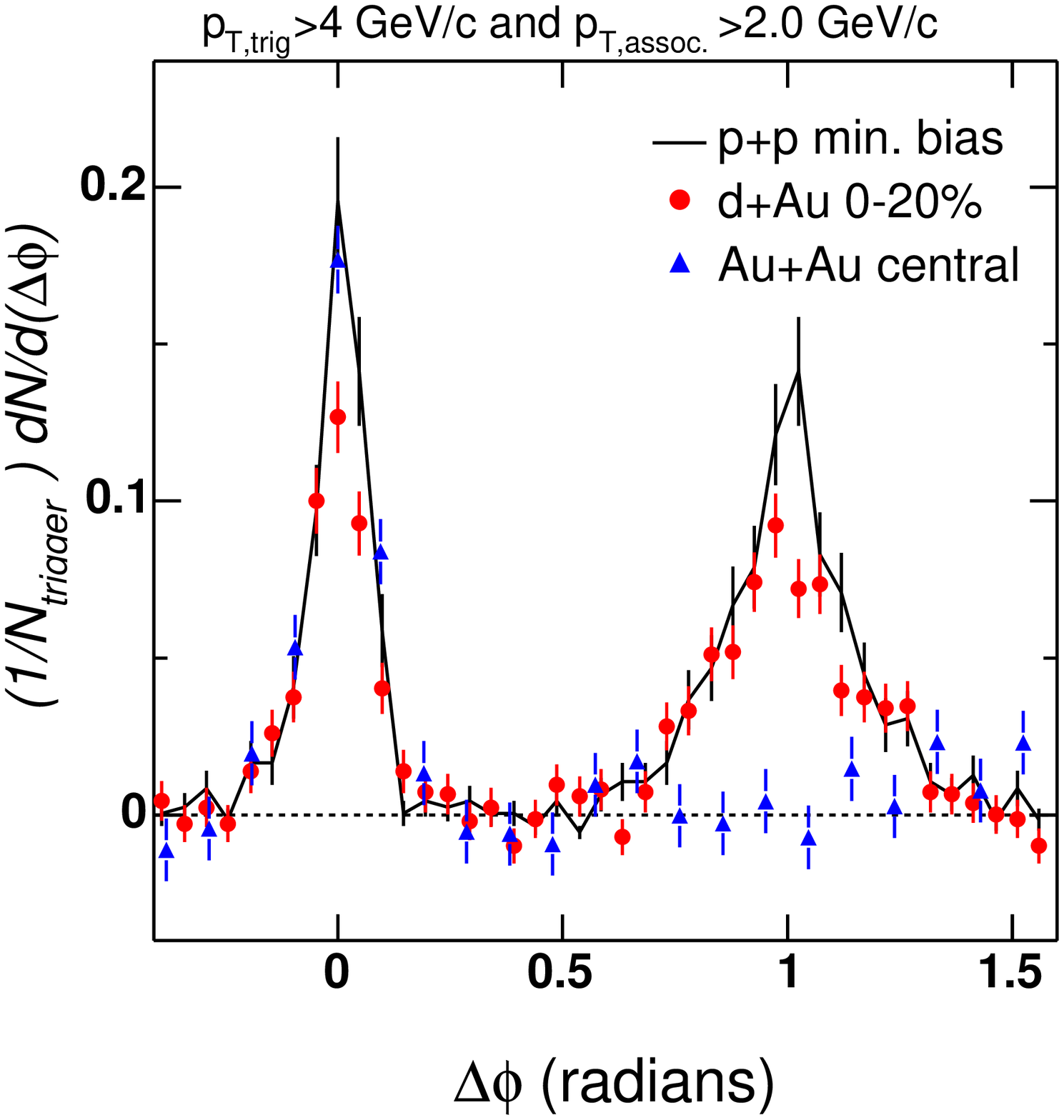}}
\vspace*{-.5cm}
\caption[] {Left: Hydrodynamic predictions~\cite{hydro} for identified
  particle $v_2$ with data from the STAR and PHENIX
  experiments~\cite{Adams:2003am,pidv2}. Center: STAR measurements of
  identified particle $R_{CP}$~\cite{Adams:2003am,msb}. Grey bands
  indicate model uncertainties in the number of collision participants
  and $\mathrm{N_{bin}}$. These uncertainties are
  correlated between species and therefore, cannot change the species
  dependence. Right: Comparison of pedestal subtracted azimuthal
  distributions (relative to high-$p_T$ trigger particles) for p+p,
  central d+Au, and central Au+Au collisions~\cite{dAu}.}
\label{figa}
\end{figure}

Consideration of the characteristic length scales associated with each
of these regions provides a simple framework for understanding the
observed dependencies in each region~\cite{ptrange}. In the {\bf
  low-$p_T$--long-wavelength limit}, observables are dominated by bulk
properties such as pressure and temperature. As demonstrated in
Fig.~\ref{figa} (left), hydrodynamic descriptions of the systems
evolution --- relying on a zero mean-free-path approximation ---
successfully describe the conversion of spatial anisotropy to momentum
anisotropy ($v_2$)~\cite{hydro}. In addition, thermodynamic models
successfully predict the relative abundances of various
particles~\cite{therm}.

In the {\bf high-$p_T$--short-wavelength limit}, hard scattered
partons first lose energy in the high density medium then hadronize in
vacuum~\cite{highpt}. As demonstrated in Fig.~\ref{figa} (right),
parton energy loss in central collisions tends to suppress away-side
jet-like correlations~\cite{dAu}. High-$p_T$ trigger particles are
predominantly generated from a fast parton escaping from the surface
of the reaction volume while the other fast parton created in the hard
scattering is directed into the reaction volume and absorbed by the
medium. Since the surviving fast partons tend to hadronize in vacuum,
the particle composition is determined by the probability of producing
a given arrangement of colorless quarks from the breaking of strings
and becomes similar to that seen in p+p collisions.

In both of the above regions the spatial distribution of the partonic
matter generated in the collision is ignored during the hadronization
process. At low-$p_T$, a locally homogenous cell of energy and
momentum radiates particles and the chemical potentials and masses of
the particles along with the matter's temperature determine the
relative particle abundances. At very high-$p_T$, after the partons
initially lose some fraction of their energy in the medium, the
medium's influence on the process of hadronization is ignored.

At {\bf intermediate-$p_T$}, however, the composite nature of the
partonic matter cannot be neglected and the low- and high-$p_T$
treatments of hadronization fail --- low-$p_T$ because it is blind to
the individuality of the matter's constituents and high-$p_T$ because
the matter is ignored entirely. The lower bound of the
intermediate-$p_T$ region can be defined by the break-down of the
hydrodynamic description of $v_2$ (an observable that is less polluted
by the post hydrodynamic/hadronic phase). The upper bound is
determined by the onset of independent fragmentation, where the
relative particle abundances return to those values measured in p+p
collisions. The lower and upper bounds were measured
experimentally~\cite{Adams:2003am} and found to be $\sim 1.5$ and
$\sim 5$ GeV/c respectively. As will be discussed below, the manner in
which the composite nature of the matter manifests itself in $v_2$ and
$R_{CP}$ provides strong evidence for the existence of {\it matter}
with {\it partonic degrees-of-freedom}.



Given the above considerations it becomes apparent that by varying the
phase-space density (with A+A, N+N, or N+A collisions) the microscopic
processes by which hadron formation occur can be studied. The manner
in which the quark-number dependencies manifest themselves at
intermediate-$p_T$ gives insight into the nature of these microscopic
processes. Particularly the larger rate-of-increase for baryon
production, {\it i.e.}  $R_{CP}^{\mathrm{Baryon}} >
R_{CP}^{\mathrm{Meson}}$, along with the quark-number-scaling for
$v_2$ shown in Fig.~\ref{figc} (left), {\it i.e.}
$v_{2}^{\mathrm{Baryon}}(\frac{3}{2}
p_{T})=\frac{3}{2}v_{2}^{\mathrm{Meson}}(p_T)$, suggests hadrons at
intermediate-$p_T$ are formed by coalescence of co-moving constituent
quarks~\cite{reco}. The quark-number dependence demonstrates that the
relevant degrees-of-freedom are partonic, while the high degree of
collectivity, developed by even the heavier strange-quarks, suggests
that the partonic degrees of freedom are locally equilibrated. We note
that the baryon--meson dependence of $v_2$ and $R_{CP}$ contradicts
expectations from simple energy loss considerations. Specifically,
while $v_{2}^\mathrm{Baryon} > v_{2}^{\mathrm{Meson}}$ would suggest
partons that fragment into baryons suffer more energy loss,
$R_{CP}^{\mathrm{Baryon}} > R_{CP}^{\mathrm{Meson}}$ (Fig.~\ref{figa})
contradicts this interpretation.

\section{Testing coalescence scaling}

\begin{figure}[htb]
\vspace*{-.4cm}
\resizebox{.5\textwidth}{!}{\includegraphics{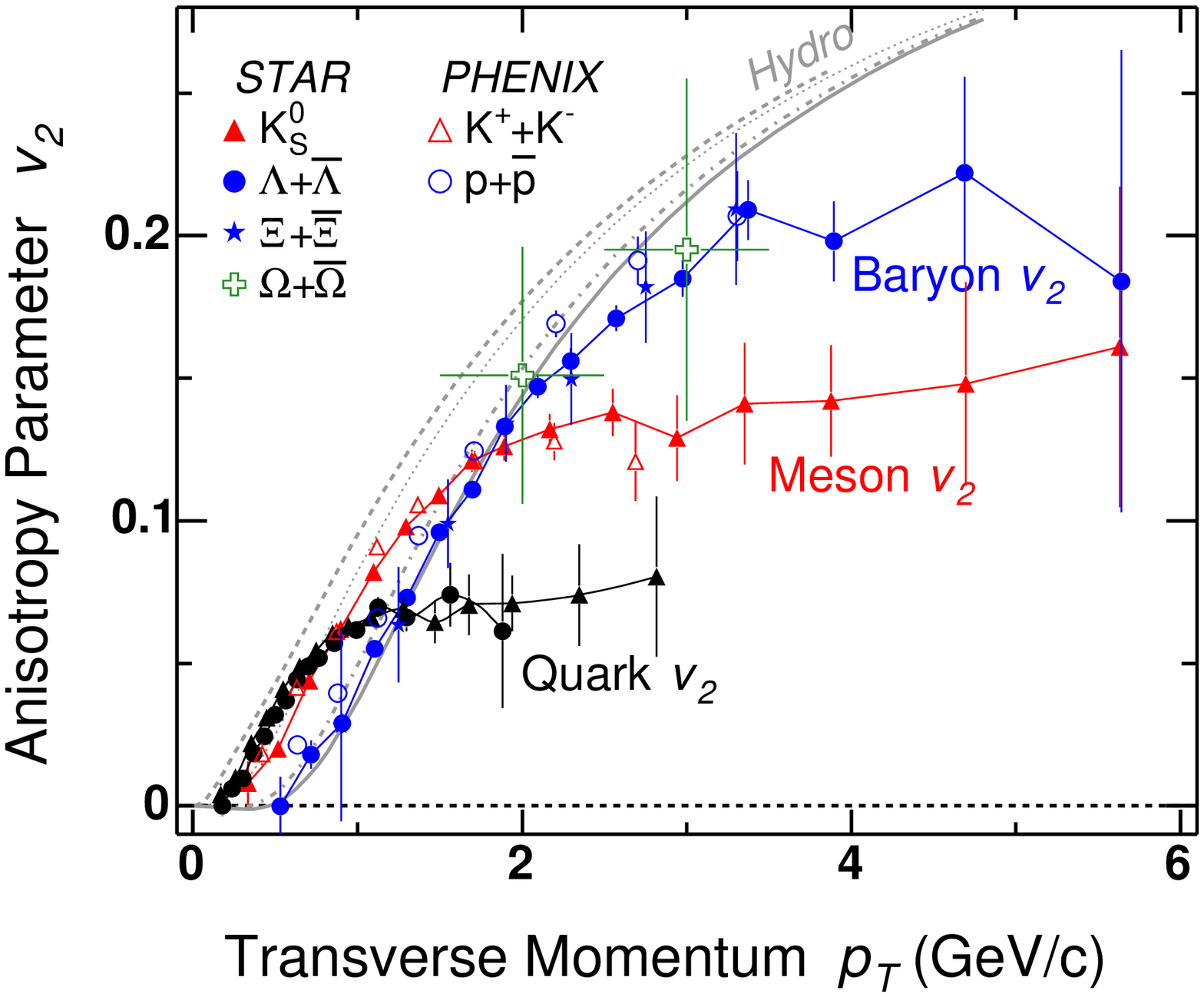}}
\resizebox{.5\textwidth}{!}{\includegraphics{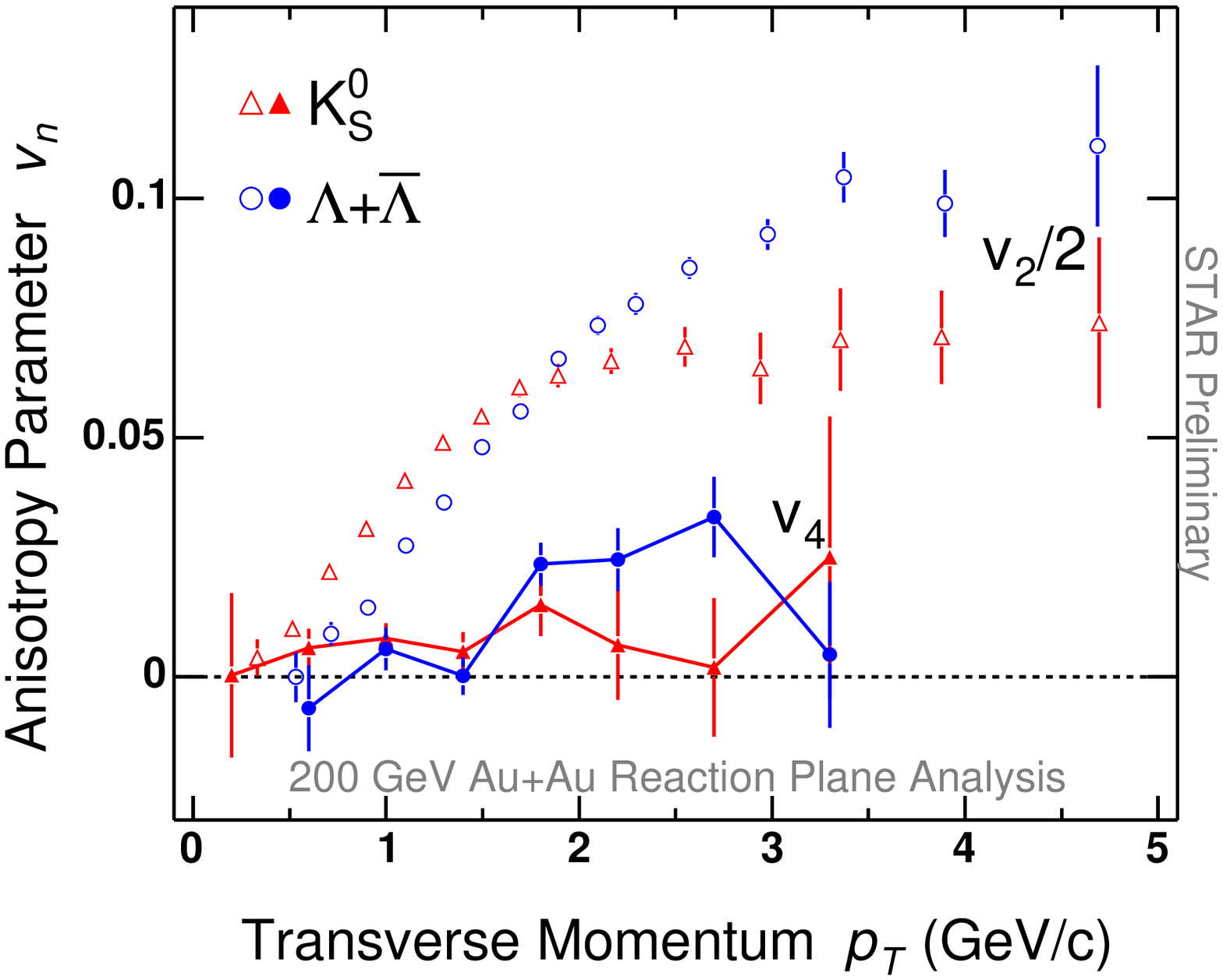}}
\vspace*{-.7cm}
\caption[]{Left: Identified particle $v_2$, hydrodynamic
  predictions and quark $v_2$ derived from NCQ-scaling. Right:
  Preliminary $K_S^0$ and $\Lambda$ $v_4$.}
\label{figb}
\end{figure}

In Fig.~\ref{figb} (left) we show $v_2$ for kaons, protons,
$\Lambda$s, $\Xi$s and $\Omega$s~\cite{Adams:2003am,msb,pidv2}. At
lower $p_T$, hydrodynamic calculations describe $v_2(p_T,mass)$ well,
with heavier particles at a given $p_T$ having smaller $v_2$
values. At $p_T>1.5$~GeV/c, the mass ordering breaks down with the
heavier baryon $v_2$ becoming larger than the lighter meson $v_2$. We
also show the quark $v_{2}$ derived by scaling the $K_S^0$ and
$\Lambda$ $v_2$ by their number of constituent quarks. In coalescence
models, the NCQ-scaled $v_2$ values reflect the anisotropy of the
parton distributions at the moment just prior to hadronization~\cite{reco}. In
Fig.~\ref{figb} (right) we show the first measurements of the higher
harmonic Fourier coefficient $v_4$ for identified particles. An
NCQ-dependence is expected to be present in higher harmonic terms ---
although in a slightly more complicated form~\cite{v4th}.
The ratio $v_4/(v_2)^2$ is preferred for model comparisons and in a
coalescence model simplifies the relationship between partonic and
hadronic anisotropies. 

\begin{figure}[htpb]
\vspace*{-.1cm}
\resizebox{.51\textwidth}{!}{\includegraphics{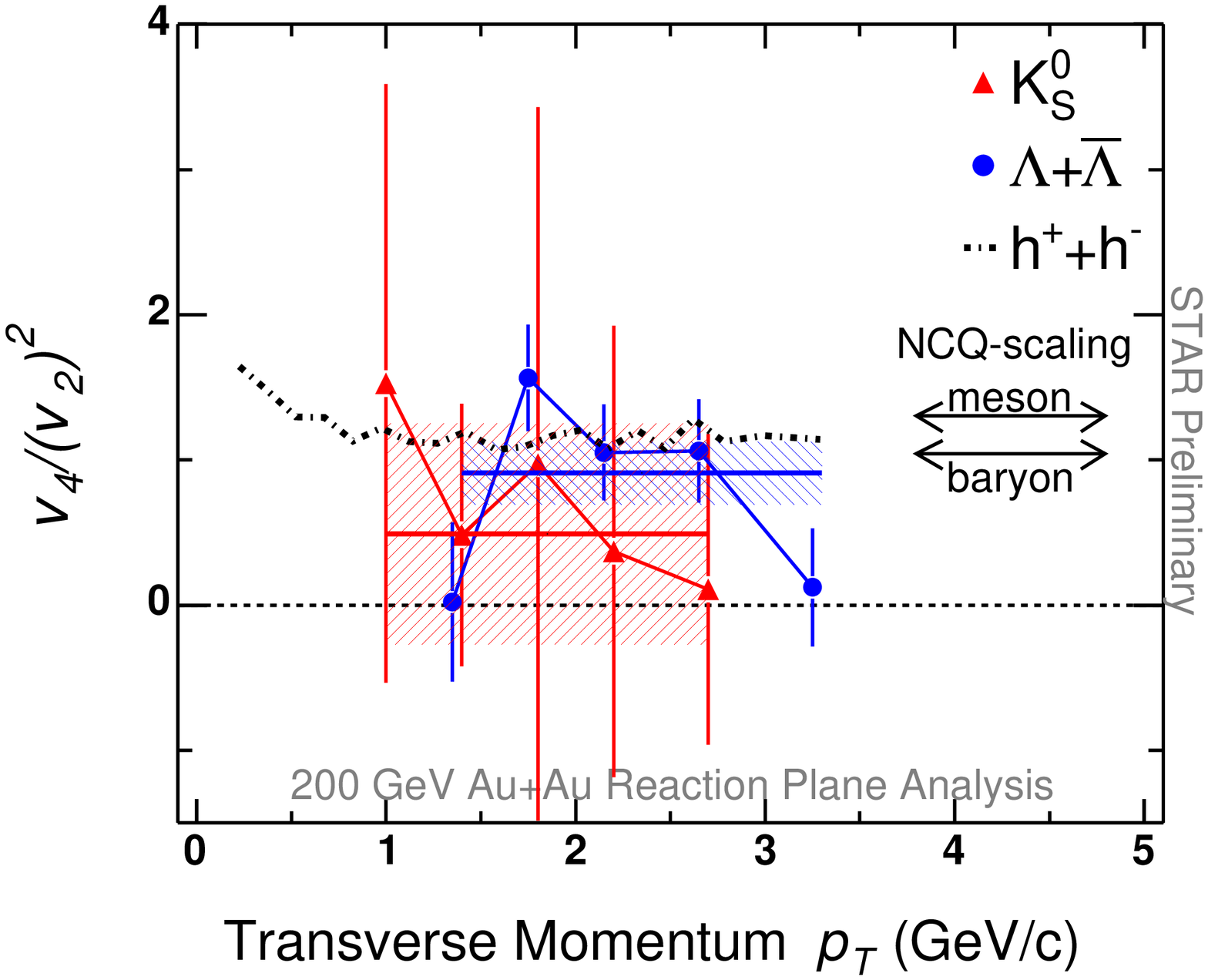}}
\resizebox{.45\textwidth}{!}{\includegraphics{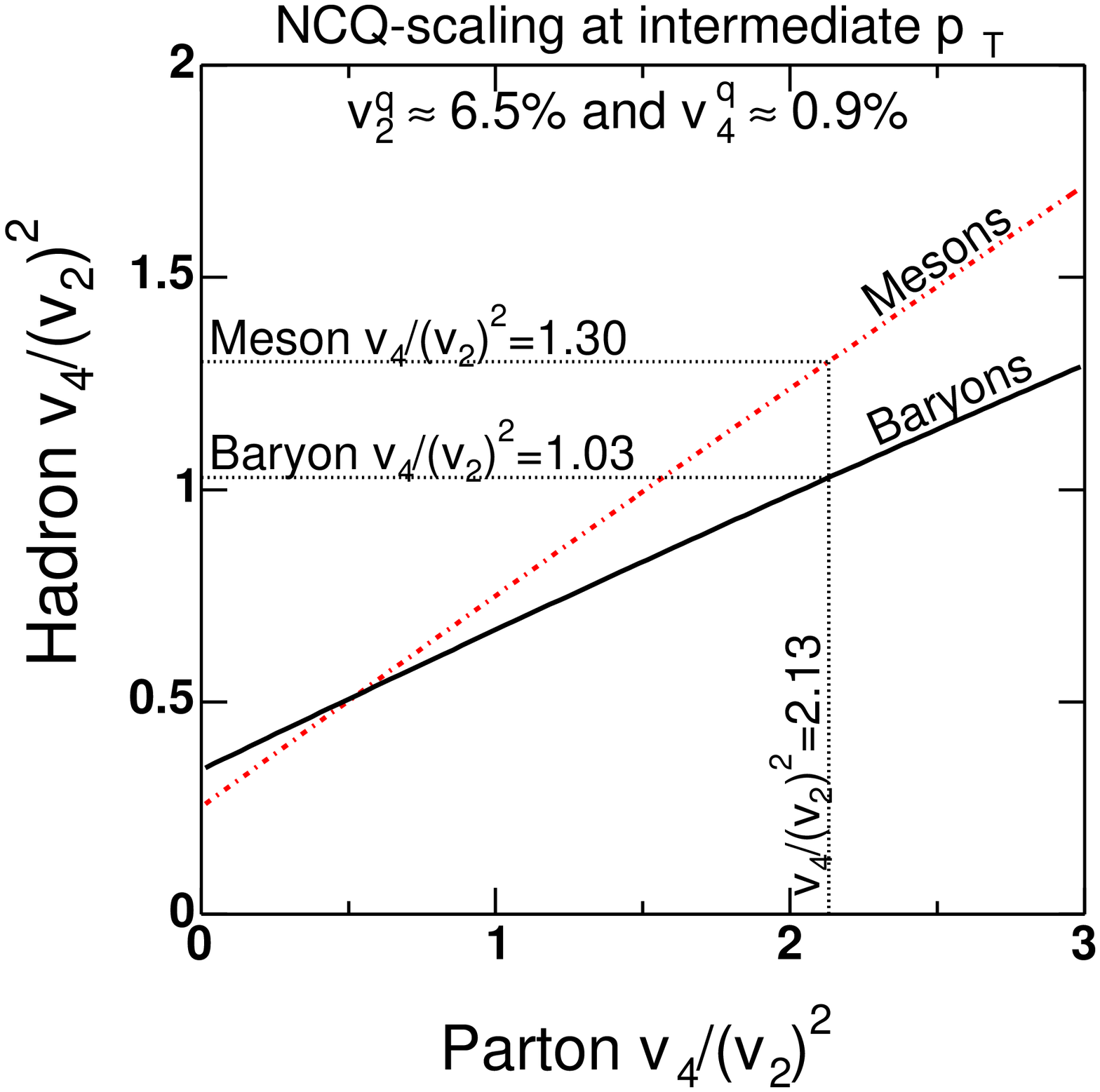}}
\vspace*{-.7cm}
\caption[] {Left: Identified particle $v_4/(v_2)^2$. Hatched regions
  indicate $\langle v_4/(v_2)^2\rangle$. Right: The relationship of
  hadron and parton $v_4/(v_2)^2$ for coalescence scaling. Values of
  $v_4/(v_2)^2$ consistant with measurements are indicated in both
  plots.}
\label{figc}
\end{figure}

Fig.~\ref{figc} (left) shows the ratio of $v_4/(v_2)^2$ for kaons,
$\Lambda$s, and charged hadrons~\cite{v4}. Although the statistical uncertainty
in $v_4/(v_2)^2$ for kaons is large, if we assume that the $\Lambda$s
are representative of all baryons we can infer that the meson ratio
must be above the inclusive charged hadron ratio. Using this
consideration, we attempt to find values for parton $v_2$ and $v_4$
that when scaled by the number of constituent quarks are consistent
with the data. Fig.~\ref{figc} (right) shows the relationship between
the partonic and hadronic anisotropies in a coalescence model~\cite{v4th}. The
specific values of parton anisotropy --- $v_2\approx 6.5\%$ and
$v_4\approx 0.9\%$ --- that are consistent with preliminary
measurements are indicated in the figures.

\section{Conclusions}\label{concl}
Identified particle $R_{CP}$ and $v_2$ measurements made in Au+Au
collisions at RHIC have revealed an apparent constituent-quark-number
dependence in the region $1.5 < p_T < 5$~GeV/c. The $v_2$ measurements
follow a quark-number scaling predicted by models of hadron formation
by coalescence of co-moving partons. The $R_{CP}$ measurements show
that baryon production increases more sharply with centrality than
meson production --- an observation that also supports a picture of
hadron formation by coalescence or recombination. Preliminary
measurements of identified particle $v_4$ have been presented and are
shown to be consistent (within the large statistical uncertainties)
with quark-number scaling predictions. Recombination or coalescence of
partons provides the most economical and compelling explanation for
these measurements. RHIC data that is currently being analyzed will
allow experimentalists to test these models more stringently. If the
initial observations of a quark-number dependence are verified, the
conclusion that a highly interacting matter with parton
degrees-of-freedom has been created at RHIC will be
unavoidable. Furthermore, a greater understanding of the hadron
formation process will have been achieved --- perhaps revealing how
quarks and gluons from the very early universe formed into the
colorless hadronic matter that dominates the visible universe today.

\section*{Acknowledgements}
The author thanks the conference organizers D. Kharzeev, R. Lacey,
M. Lisa, and S. Panitkin along with V. Greco, A. Tang, and S. Gulerce
for stimulating discussions.

\vfill\eject
\end{document}